\documentclass{article}

\usepackage{PRIMEarxiv}

\usepackage[utf8]{inputenc} 
\usepackage[T1]{fontenc}    
\usepackage{hyperref}       
\usepackage{url}            
\usepackage{booktabs}       
\usepackage{amsfonts}       
\usepackage{nicefrac}       
\usepackage{microtype}      
\usepackage{lipsum}
\usepackage{fancyhdr}       
\usepackage{graphicx}       
\graphicspath{{media/}}     
\usepackage{subfigure}
\title{Say Your Reason: Extract Contextual Rules In Situ for Context-aware Service Recommendation}

\author{
  \textbf{Yuxuan Li}\\
  Tsinghua University \\
  Beijing, China\\
  \texttt{yuxuan-l20@mails.tsinghua.edu.cn} \\
      \AND
\textbf{JiaHui Li}\\
Tsinghua University\\
Beijing, China\\
\texttt{ljh1304607285@hotmail.com}
  \And
  Lihang Pan\\
  Tsinghua University \\
  Beijing, China\\
  \texttt{plh18@mails.tsinghua.edu.cn} \\
  \And
  Chun Yu\thanks{Indicates the corresponding author.}\\
  Tsinghua University \\
  Beijing, China\\
  \texttt{chunyu@mail.tsinghua.edu.cn} \\
  \And
  Yuanchun Shi\\
  Tsinghua University \\
  Beijing, China\\
  \texttt{shiyc@tsinghua.edu.cn} \\
}

\begin{document}
\maketitle

\begin{abstract}
This paper introduces SayRea, an interactive system that facilitates the extraction of contextual rules for personalized context-aware service recommendations in mobile scenarios. The system monitors a user’s execution of registered services on their smartphones (via accessibility service) and proactively requests a single-sentence reason from the user. By utilizing a Large Language Model (LLM), SayRea parses the reason and predicts contextual relationships between the observed service and potential contexts (such as setting the alarm clock deep in the evening). In this way, SayRea can significantly reduce the cognitive load on users in anticipating future needs and selecting contextual attributes. A 10-day field study involving 20 participants showed that SayRea accumulated an average of 62.4 rules per user and successfully recommended 45\% of service usage. The participants provided positive feedback on the system's usability, interpretability, and controllability. The findings highlight SayRea's effectiveness in personalized service recommendations and its potential to enhance user experience in mobile scenarios.
\end{abstract}


\begin{figure}[ht]
  \includegraphics[width=\textwidth]{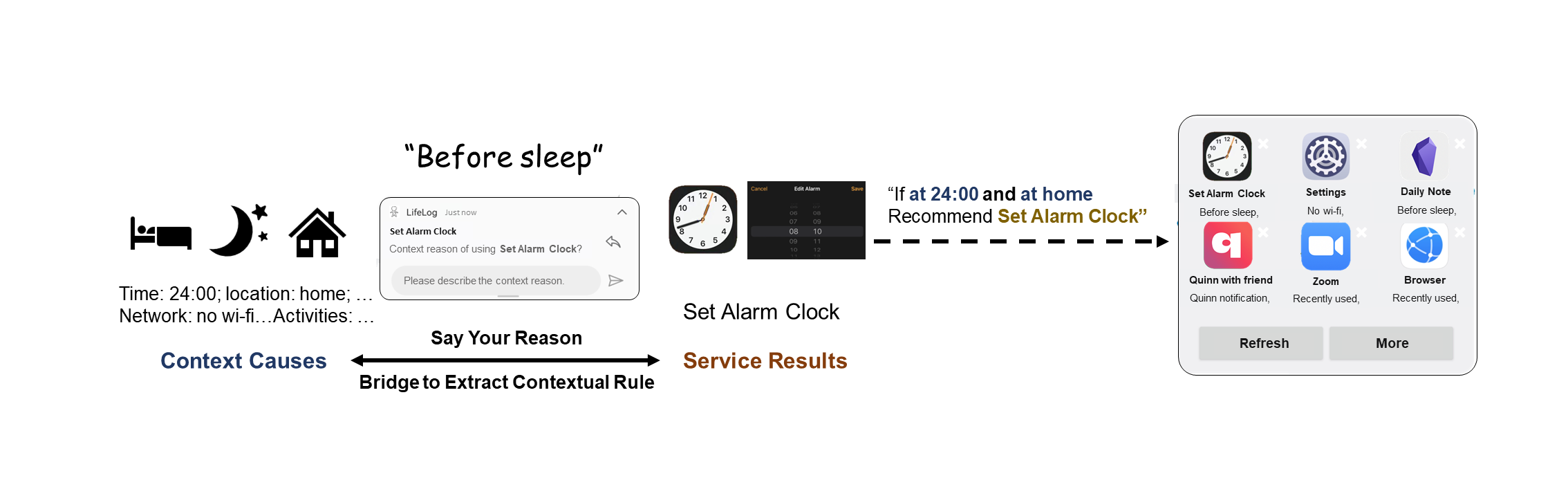}
  \caption{When user use services on their smartphones, SayRea proactively requests user to say the reason in situ. SayRea use the reason as bridge between context causes and service results to extract contextual rule for context-basd service recommendation. Next time if the context meets rules, our system provide quick access for the services on home screen and lock screen.}
  %
  \label{fig:teaser}
\end{figure}

\section{Introduction}

With the increasing number of mobile applications and in-app services, accessing services occupies more and more users' attention, sometimes even surpassing the services themselves\cite{Ma2020Roaming}. Context-aware recommendation enables efficient and comfortable user access to smartphone services. Contexts (e.g., deep in the evening) and services (e.g., setting the alarm clock) have strong relationships. Existing systems often utilize contextual rules (with the context as the cause and the service as the effect) that can directly describe these relationships (such as "if it's late at night and about to sleep, then recommend the alarm clock service"). Introducing end-user to edit contextual rules meets users' requirements for quick service accessing and offers interpretability and controllability. However, existing user rule editing systems\cite{Steven2014IFTTT,Ilarri2022An} have high entry barriers. Users have a heavy operation load on selecting context attributes and a significant cognitive load on anticipating future needs to build rules deliberately in advance.

In order to solve the high entry barrier problem, this paper proposes SayRea (Say your Reason), an interactive system that facilitates the extraction of contextual rules for personalized context-aware service recommendations in mobile scenarios using a Large Language Model(LLM). The rule extraction method in SayRea has two characteristics: simple interaction mechanism and appropriate interaction timing. (1) SayRea's interaction mechanism requests users of single-sentence natural language reasons for services instead of complete and accurate information about the contexts and services. The users do not need to manually select any context attributes, SayRea does the job for them. (2) SayRea extracts rules in situ by interacting with users for automatically selecting essential contexts when the usage of services occurs. In situ, users do not need to deliberately anticipate future needs, and the system can automatically perceive necessary surrounding contexts.

SayRea utilizes a pre-trained large language model(LLM)\cite{Brown2020Language} to fully leverage the semantic information in users' reasons, in order to automatically identify essential context attributes and bridge the contexts and services when building contextual rules. 
In most cases, SayRea identifies essential context attributes using only two components: (1) semantic translations of contexts and services in the form of natural language representations; (2) users' reasons. It can then extract contextual rules based on the identified context attributes. In some cases, as contextual rules accumulate over time, SayRea is able to identify the existing contextual rules that match the current context attributes. This allows users to engage in interactive rule extraction without even being required to provide a natural language reason, thereby further reducing the burden on users.

SayRea also enables users to delete extracted contextual rules when false or expired rules accumulate. SayRea uses a tree structure to manage rules and ranks recommendations when multiple rules satisfy the contexts after the accumulation.

We implemented SayRea on the Android platform. SayRea collects and labels commonly used services in different applications and monitors the services on users' smartphones with the Accessibility Service. It applies context awareness\cite{Musumba2013Context-awareness} to collect the contexts and semantically translate them using feature engineering. We also designed interactive interfaces for rule extraction and service recommendation.

A 10-day field study with 20 participants was conducted to evaluate the SayRea system. SayRea extracted an average of 62.4 rules from each user, indicating the high usability of the extraction method. 
The service coverage(a service being used is covered if it is also being recommended) increased over time and reached a maximum of 45\%, demonstrating the system's recommendation capability. Users' subjective feedback showed a high acceptance of SayRea, and users thought that the system had good usability, interpretability, and controllability.

In summary, the main contributions of this paper are:
\begin{itemize}
\item The proposal of an in situ interactive method for extracting contextual rules. This method enables low-burden user interaction at appropriate times to extract contextual rules between contexts and services using a Large Language Model.

\item The implementation of a context-aware service recommender system with contextual rule extraction. The system is interpretable, controllable, and able to personalize.
\end{itemize}

In the following sections, we first summarize related work on context-aware service recommender systems, rule editing. In Section 3, we introduce the method for extracting contextual rules. Then we elaborate on the context-aware service recommendation in Section 4. Section 5 is the implementation of SayRea's system modules. And section 6 presents the field study of SayRea and its results. Finally, we provide discussions, limitations, and a conclusion for this work.

\section{Related Work}
This section will summarize the previous work on context-aware service recommender systems and rule editing. We mainly introduce the contributions of the SayRea system presented in this paper compared to these related works.

\subsection{Context-aware Service Recommender System}
Context-aware service recommendations\cite{María2021context-aware, Norha2018Characterizing} can optimize users' smartphone service usage processes. Currently, commercial application recommendations on smartphones, such as recommendations in Siri Suggestion\cite{Carrasco2019Siri}, and some research work in the field, like DeepApp\cite{shen2019deepapp}, CMARA\cite{Zhu2021CMARA}, generally, adopt end-to-end deep learning models for recommendations. These models are intelligent but have issues in interpretability\cite{Meng2022Survey, Yongfeng2020Explainable} and controllability\cite{Tsai2021controllability,Jannach2017Control}, leading to lower usage rates. Although some context-aware recommender systems attempt to provide explanations for recommender models\cite{Scheel2014Explanations}, such as Amit\cite{Amit2022Evolving} and Zhang\cite{Zhang2019Context-Aware}'s work, and CAESAR\cite{Li2021CAESAR}, they still require a large amount of data and time to accumulate data.

The LLA application\cite{Karchoud2017Long-life} introduces context management and service recommendations based on specific algorithms with relatively fewer contexts. These systems show a certain degree of interpretability and can adapt to user usage. Compared to SayRea, however, these systems are more complex to manage and still lack controllability. Context-aware service recommendations can also be sourced from association rule mining\cite{Zhang2002Causality} under the guidance of machine learning methods\cite{Shaina2019overview}, such as ABC-ruleminer\cite{Iqbal2020ABC-RuleMiner}. Recommender systems based on such association rules will have better interpretability, but still have data dependency, potential redundancy of rules\cite{Zaki2004Mining}, and lack of controllability.

SayRea interacts with users to extract causal rules for context-based service recommendations, requiring minimal data support. It has advantages in a mobile intelligent system environment where users value data privacy while achieving low implementation costs, interpretability, and controllability.

\subsection{Rule Editing}

The ability for end-users to edit rules is the foundation of personalization and controllability in SayRea. There are related work in the field of human-computer interaction. Rule editing is common in task automation systems\cite{Coronado2016Automation}, such as IFTTT (If This, Then That)\cite{Mi2017IFTTT}, which allows defining actions by connecting applications to complete tasks and actions automatically. According to IFTTT's website, more than 600 applications and devices collaborate with IFTTT, such as Twitter, Telegram, Google Drive, and Alexa. Similarly, in the Shortcuts on iOS platforms\cite{Shortcutonline}, specific conditions can be defined under which a shortcut should automatically run. Users need to choose from a list of many trigger conditions.

Rule editing through a "form-filling" method is considered very tedious and boring\cite{Karchoud2019One}. Some research has proposed systems like SECE (Sense Everything, Control Everything)\cite{Boyaci2011Bridging} and DOTT (Do This on That)\cite{Chihani2013DOTT}, which allow users to edit rules using natural language, and "one app to rule it all"\cite{Karchoud2019One} enables users to edit rules through methods such as drag-and-drop. These works focus on optimizing the methods for user rule editing but still rely on a priori rule editing and require formalized expressions, which can lead to cognitive and interaction burdens. The timing of in situ interactions has been discussed in feedback procedures for recommender systems, like AppEcho\cite{Seyff2014AppEcho}. SayRea uses in situ interaction timing to allow users to provide simple reasons, resulting in very low burden rule editing in terms of interaction and cognition.

\subsection{In situ interaction}
An important challenge in user intervention in context-aware recommendation systems (e.g., editing rules) lies in the timing of interactions. Context is constantly changing, leading to a significant gap between the context in which users need a service and the context in which they interact with the system \cite{Karchoud2017Long-life}. This introduces an additional mental burden: users must recall or envision a non-existent scenario. This gap also makes it difficult for users to get immediate result on their interventions, which not only reduces their confidence but also diminishes their willingness to interact.

Sayrea applies in situ interaction to address the problem mentioned above. In situ interaction meaning that the interaction takes place where and when the target context exists. It has been widely used in visualization systems \cite{korn2013potentials}. For examlpe, reaseachers tried to augment code with in situ visualizations to help program understanding \cite{hoffswell2018augmenting}. However, this method has not been applied to recommendation systems yet.

\section{Contextual Rule Extraction}
This section mainly introduces the contextual rules for context-based service recommendation in SayRea and its in situ interactive extraction method.

\subsection{Contextual Rules of Contexts and Services}
Contextual information is often multi-dimensional, involving aspects like time, location, and activity status. Each dimension contains various features. For instance, time could have features such as the day of the week and the specific hour. At any given moment, each feature has a unique value, such as [Monday, 7 a.m.]. We call these pairs of features and values "context attributes." These attributes provide LLMs with the necessary information to identify the most relevant attributes for making service recommendations. The results in the contextual rules are services, which include apps and in-app services on mobile phones.

Context-aware service recommendations are based on these contextual rules, expressed as "if a certain context occurs, recommend a certain service." The context cause is a combination of multiple context attributes, describing complex contexts. The formalized expression is as follows:  $$if \; A_1 \; and \; A_2 \; and \; A_3 \; ... \; then \; recommend \; S_A.$$For example,  in contextual rule "if the user is in the dormitory at 24:00 at night and lying down, recommend the alarm clock service", $A_1$ represents being in the dormitory, $A_2$ represents that it is 24:00 at night, $A_3$ represents lying down, and $S_A$ represents the alarm clock.

\subsection{Contextual Rule Extraction}

\begin{figure}[h]
  \centering
  \includegraphics[width=\linewidth]{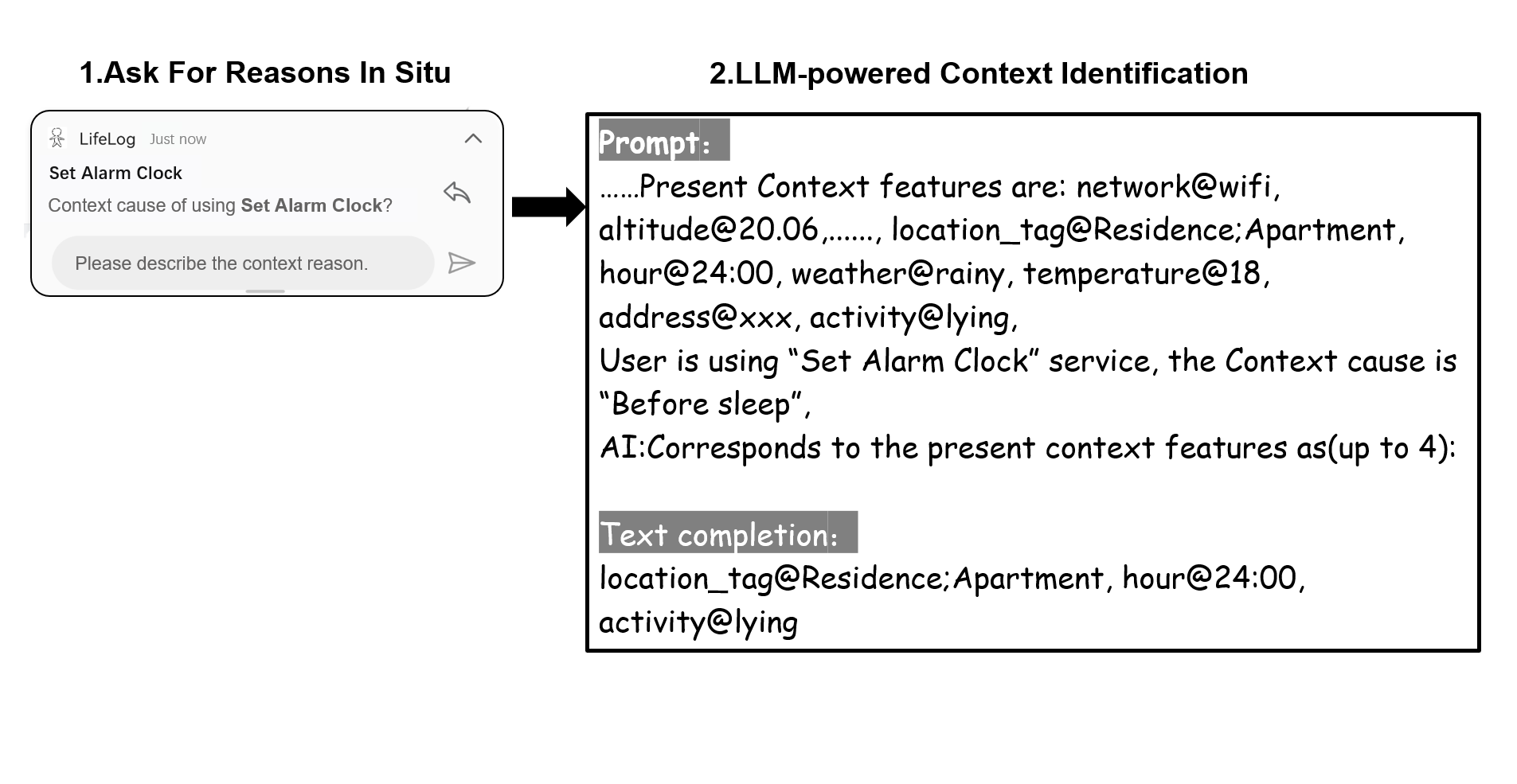}
  \caption{An example of contextual rule extraction through two steps.}
  \label{fig:extraction_context_rule}
\end{figure}

 SayRea uses proactive interaction requests in situ to allow users to establish contextual rules by saying the reasons in natural language, reducing the burden of users and lowering the entry barrier for user rule editing. The overall process is shown in the figure\ref{fig:extraction_context_rule}. 
SayRea first requests interaction with users for them to say natural language reasons. The system then uses an LLM to predict identify essential context attributes. The cause (a set of identified context attributes) and the result (the service) then form a contextual rule,

\subsubsection{Ask For Reasons In Situ}
When SayRea recognizes(further elaborate in \ref{sec:Service Recognization Module}) that the user is using a service, it initiates an interactive request to ask the user for a reason. 

The main content of the interactive request is to ask the user to provide a natural language reason for using the service. A simple reason is all that the user need, without attention for expressing the context details or using formatted sentences.

This in situ(service recognized) timing can reduce the burden of interaction: 1) The operational burden is insignificant: the system perceives current contexts and services, so the user only needs to say the reason. 2) The cognitive burden is negligible: the user is in the current context and using a needed service, not requiring anticipation of future needs. 

\subsubsection{LLM-powered Context Attributes Identification}

\begin{figure}[ht]
  \centering
  \includegraphics[width=0.8\linewidth]{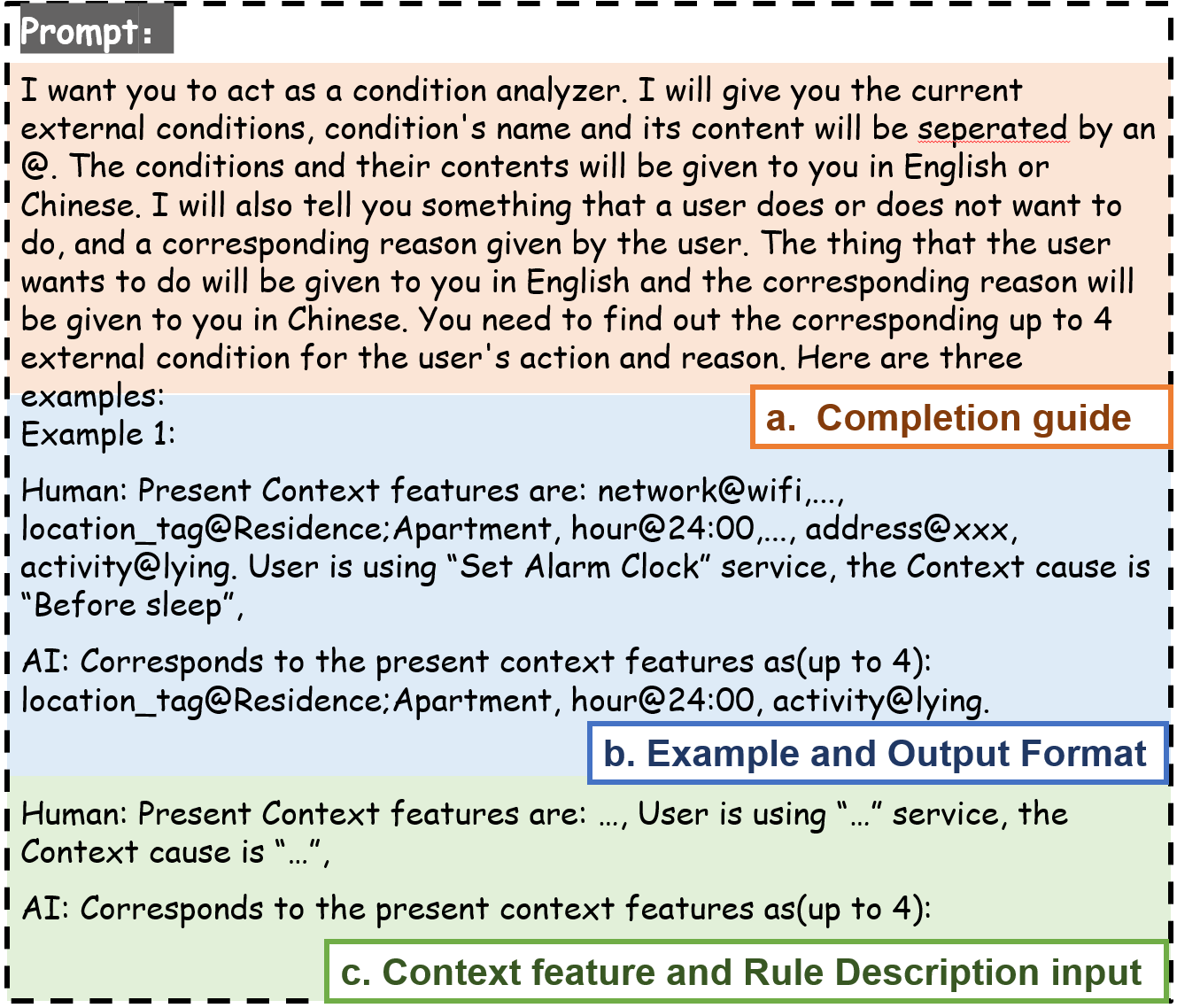}
  \caption{Prompt design for LLM-powered context cause extraction with its purpose decomposition.}
  \label{fig:prompt}
\end{figure}

LLMs are able to map the open-ended natural language reason to the corresponding context attributes using general knowledge (e.g., the natural language reason "very hot" corresponds to the context attribute "temperature: 28 degrees Celsius"). Identifying essential context attributes can be transformed into a text completion task. The input is the user's natural language context reason, the system's perception of the current service, and the semantic representations of all context attributes. The output is a list of predicted context attributes, and their combination represents the context cause in the contextual rule. To better accomodate users' natural language reasons, we choose a dialogue-form prompt\cite{cantino2021prompt} design based on small-scale trial with LLMs. Figure \ref{fig:prompt} shows the prompt and its purpose decomposition, including a) guidance for text completion; b) an example of the completion format, which outputs a list of context attributes; c) the current natural language reason of the context reasons and input of the context attributes.

Since the task has been transformed into a text completion task, any LLM with text completion capability can be used, not limited to GPT--3.5\cite{Brown2020Language} (which we used). Accurate identification of the essential context attributes is important to the correct accumulation of contextual rules and performance of the recommender system. We have validated the LLM's ability in this task through further evaluation (see Section 7).

\subsection{Accumulation and Optimization}
\label{sec:Accumulation and Optimization}
After a contextual rule is extracted, the user's single-sentence reason is accumulated along with the contextual rule in the system. During another contextual rule extraction process, if SayRea detects that the current context attributes meets an accumulated contextual rule's cause, it directly asks whether a contextual rule can be extracted based on the single-sentence reason corresponding to the context cause, without requesting the user to give the reason. Such as the "\textit{Home time} and \textit{Reading finished}" in Figure\ref{fig:request & confirm}(a), shown as predicted reasons on the user interface. The contextual rule extraction process is optimized by reducing the interaction steps with accumulated context causes and reducing interaction burdens.

\section{Service Recommendation}
SayRea recommends services by generating a list of services and recommendation reasons based on the current context, accepts quickly accessing services and feedbacks for incorrect recommendations from users.

SayRea has the following characteristics for service recommendation:
\textbf{Low cost}: SayRea mainly uses contextual rule-based recommendation, saving time and professional personnel for data collection and model training.
\textbf{Interpretability}: SayRea semantically analyzes all the contexts and services involved in the process, and the recommendation based on contextual rules has explicitly recommendation reasons.
\textbf{Personalization and controllability}: the contextual rules in SayRea origin from user editing, and users can manually modify the rules based on their specific needs. 

\subsection{Cold-start Recommendation}
Cold-start recommendations consist of a list of recommended services in the order of the user's most recent use. This is similar to the existing service recommender systems\cite{Carrasco2019Siri} on smartphones.
\subsection{Contextual Rule-based Recommendation}
SayRea acquires all the current context attributes and recommends the corresponding result services when the contextual rules are met.
\subsection{Recommendation Feedback}
The user can give feedback for incorrect recommendations on the recommended results of the SayRea system. SayRea system initiates the contextual rule extraction again when recommendations are rejected. This time the contextual rule differs from the positive contextual rules by being negative: $$if \; A_1 \; and \; A_2 \; and \; A_3 \; ... \; then \; not \; recommend \; S_A.$$
\subsection{Accumulation and Optimization}
SayRea accumulates the contextual rules in a context rule tree, which is used for optimizing the service recommendation process.
\subsubsection{Rule Management: Context Rule Tree}

\begin{figure}[ht]
  \centering
  \includegraphics[width=\linewidth]{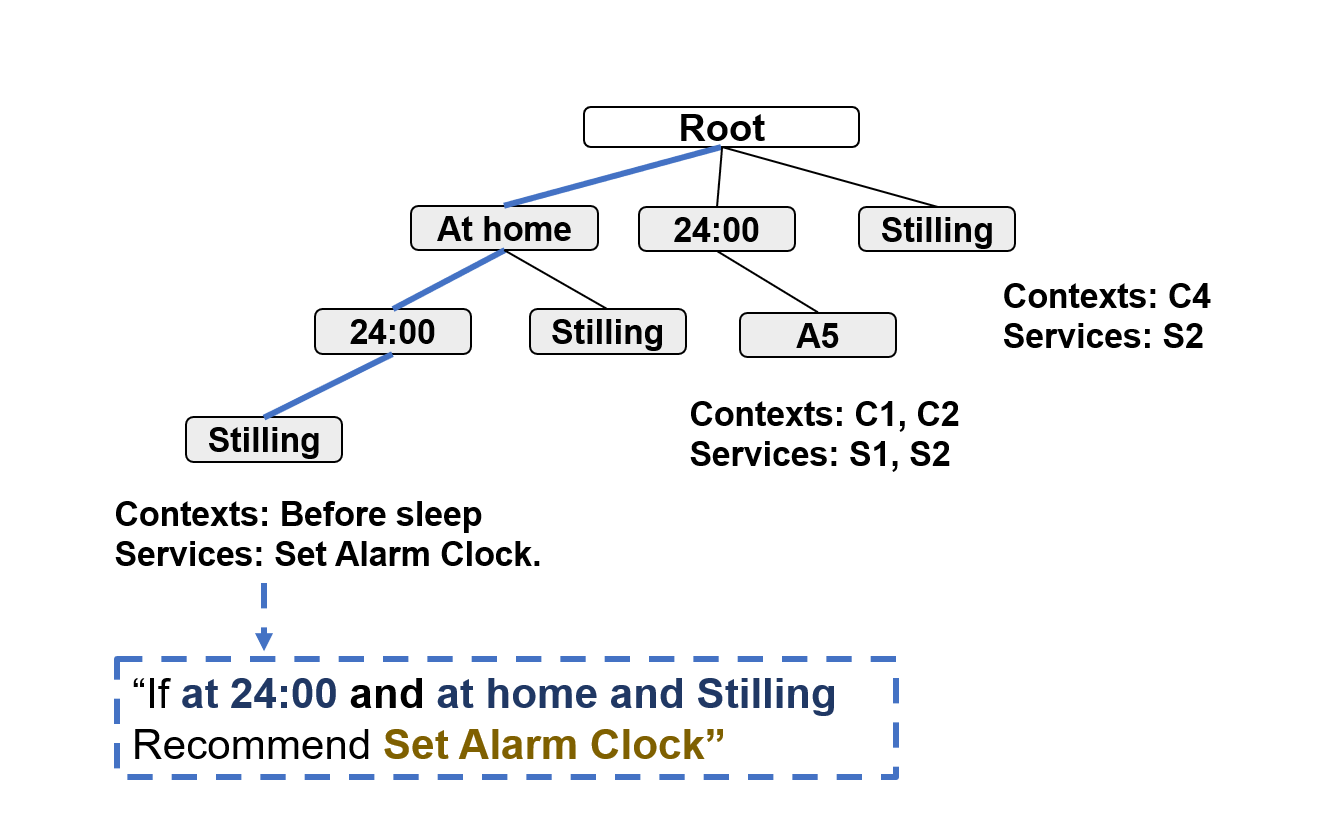}
  \caption{A context rule tree Example. Nodes of the tree represent context attributes(At home, 24:00, Stilling, A5), and some nodes also store the context reasons(Before sleep, C1, C2, C4) and service results(Set Alarm Clock, S1, S2) for contextual rules.}
  \label{fig:ruletree}
\end{figure}

Wu's item recommendation rule tree \cite{Wu2020Tree} inspired our context rule tree. As shown in Figure \ref{fig:ruletree}, the nodes of a context rule tree represent context attributes, and the path from the root node to a node is a combination of context attributes. If there is a service at a node, it means that this node corresponds to a contextual rule: if the combination of context attributes of the node is satisfied, the corresponding service is recommended. If the node has a context reason, this context reason is a natural language representation of the node's context attributes combination. Context rule tree can store accumulated different contextual rules for a same service or a same contextual rule for different services. To prevent redundancy, if two contextual rules share the same combination of context attributes, only one branch will be stored in the tree because we enforce context attributes of the parent node superior than those of the child node. Using such a tree structure can cope with the situation of sparse context attributes but complex context feature combinations, save storage space, and perform queries quickly.

\subsubsection{Context Rule Tree-Based Recommendation}
The context rule tree can find all paths that satisfied the context attributes of each node by the current context, and so acquire all services that match the contextual rules in the current context. The number of occurrences of the services and the depth of the tree correspond to the degree of matching with the current context and the complexity of the context cause, respectively. When recommending services, the services can be ranked according to the number of occurrences and the depth of the tree.

For the negative contextual rules extracted based on users' feedback, SayRea builds a negative context rule tree. For the current context, services that are acquired from the negative context tree are removed from the service recommendation list.

\begin{table*}[ht]
\centering
  \caption{Context in SayRea system}
  \label{tab:context}
  \begin{tabular}{ccc}
    \toprule
    Type&Awareness&Features\\
    \midrule
    Time & Timestamp & day of the week, time period, o'clock\\
    Network & Network manager & network type, SSID, speed\\
    Bluetooth & Bluetooth sensor & Bluetooth state, connected counts, connected devices\\
    Weather & Third party API & weather type, temperature\\
    Location & GPS, Third party API & altitude, address, location tag, position of interest(POIs)\\
    Activities & Activities recognization & on foot/running/walking/in vehicle...\\
    Service & Mobile phone logs & recently used app, unlock screen time\\
    Notification & Notification manager & recent notification, current cnotifications\\
  \bottomrule
\end{tabular}
\end{table*}

\section{Implementation}
\begin{figure}[ht]
  \centering
  \includegraphics[width=\linewidth]{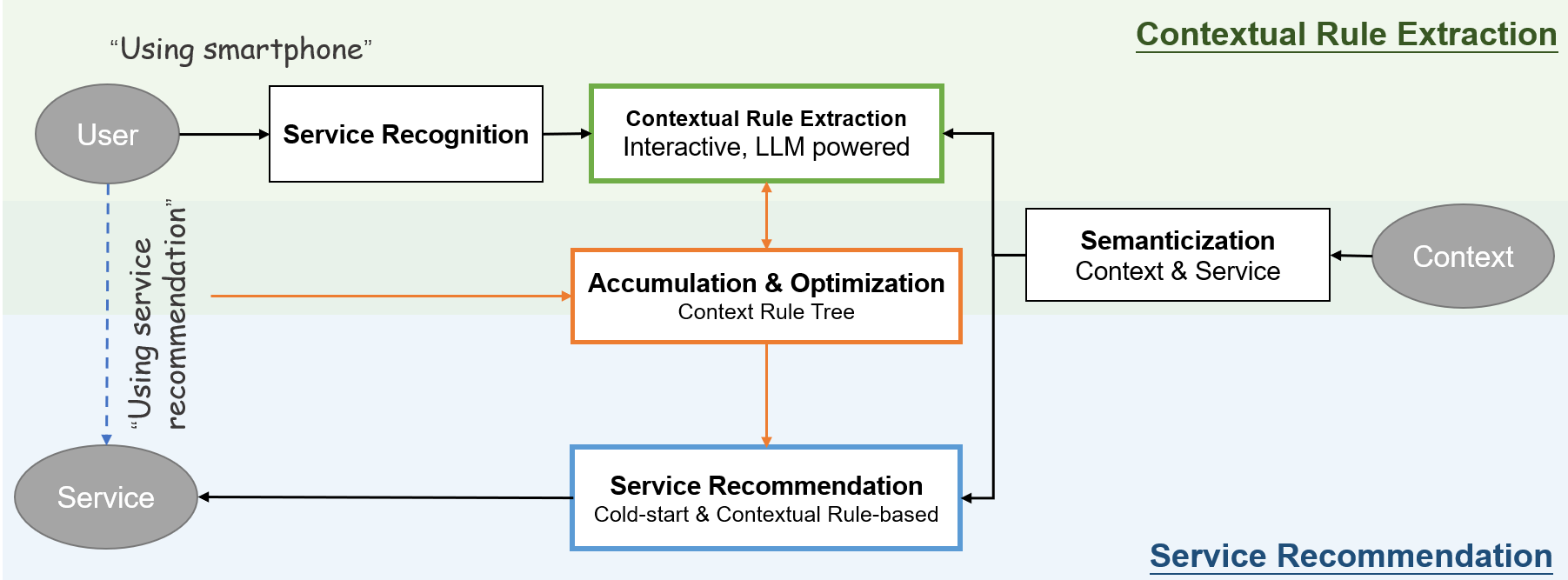}
  \caption{SayRea's system flow. SayRea extracts contextual rules while users using smartphones, and supports using service recommendations.}
  \label{fig:systemflow}
\end{figure}

SayRea's basic framework is shown in Figure \ref{fig:systemflow}. The system consists of two important processes: the contextual rule extraction process and the service recommendation process. The core contextual rule extraction module and service recommendation module, as well as their corresponding accumulation modules, have been described in the previous section. SayRea also implements a semanticization module for contexts and services, and a service recognition module to support those two processes. This section describes the design and implementation of these two modules and the system's user interface.

\subsection{Semanticization Module for Service and Context}
Semanticization modules are designed for interpretable interaction with users, and for better inputs of the LLM to take advantage of general knowledge.

\subsubsection{Service Semantic}
SayRea supports 41 selectively labeled in-app services from 14 different types of applications, and an unlimited number of "launch an application" type services (i.e., opening applications installed on the user's phone, e.g., "open Clock"). We provide a semantic label for each service (e.g., "set alarm clock"), and in particular, the semantic of an "open an app" service is defined as the name of the application.

\subsubsection{Context Semantic}
SayRea uses multiple contexts for service recommendation, as shown in Table \ref{tab:context}. 
Our system obtained the context data through context awareness and multiple context dimensions through feature engineering. The whole context at any moment has a value taken in each context dimension, which can be regarded as the set of context attributes. Templated translations("temperature with a value of x can be translated to \textit{x degrees}") of the context attributes yields the context semantics (such as Monday, 10 degrees, etc.). Any new context dimensions can be easily translated into contextual semantics for use in SayRea, showing the system's scalability.

\subsection{Service Recognition Module}
\label{sec:Service Recognization Module}
\begin{figure}[ht]
  \centering
  \includegraphics[width=\linewidth]{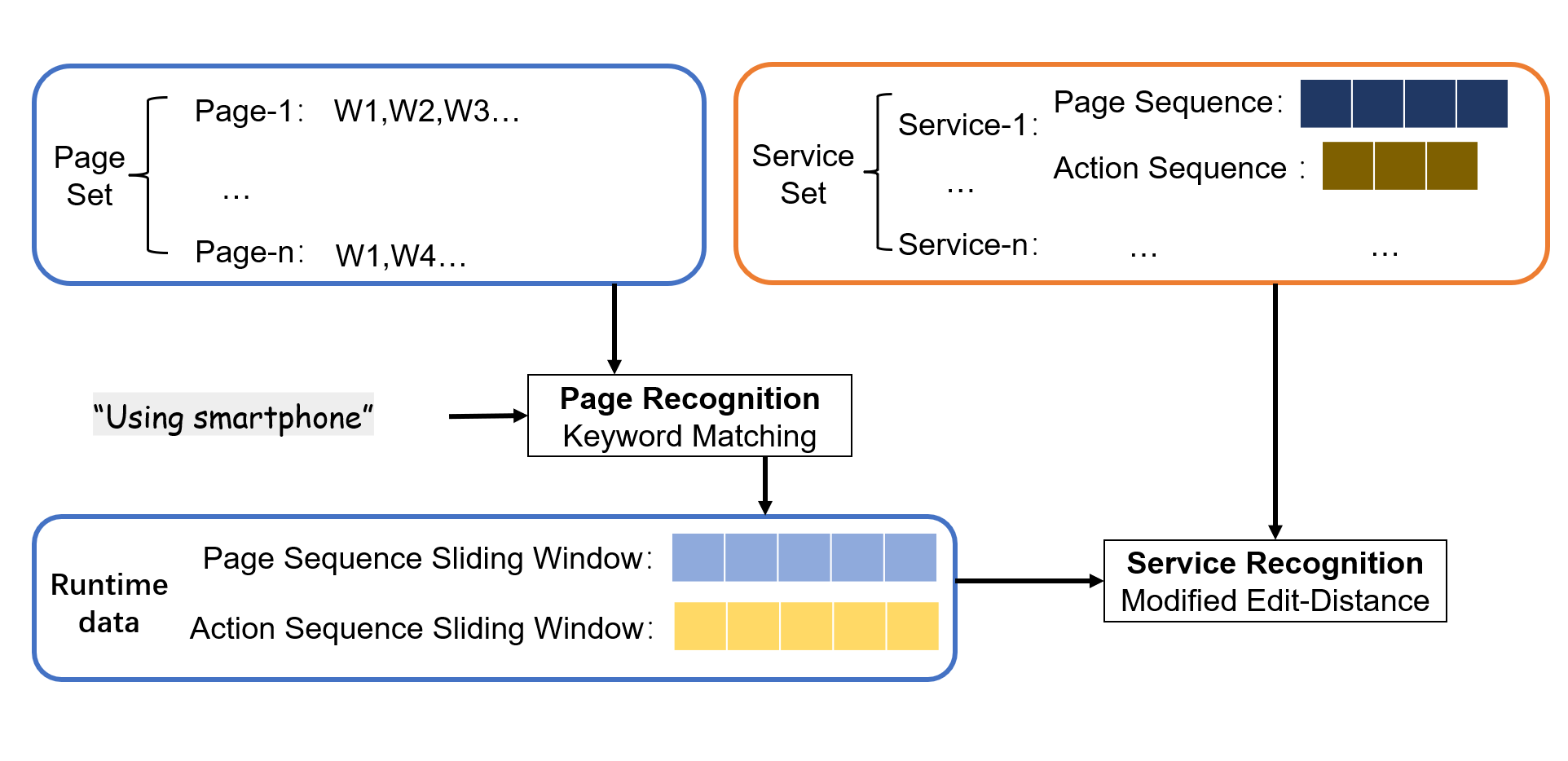}
  \caption{Service recognition produce.}
  \label{fig:serviceRecognize}
\end{figure}

 This module ensures an in situ interaction timing for the contextual rule extraction method when users are using services. 
 SayRea's service recognition process for labeled in-app services is shown in Figure \ref{fig:serviceRecognize}. First, the system labeled a predefined set of services with the texts from the service operation sequences and the keywords (static text or content descriptions\cite{ContentDescription} on the page) in the page sequence.

 We use the Accessibility Service\cite{AccessibilityService} to obtain the pages and actions running on the phone. Page recognition is performed by keyword matching (if the run-time page's 80\% keywords match the labeled page's keywords), and sliding windows for both pages and actions are maintained separately. When a page is recognized SayRea uses a modified shortest edit distance algorithm(Levenshtein distance\cite{Levenshtein}, the edit distance for deleting a sliding window sequence is set to 0 to avoid a large sliding window sequence causing a large edit distance) to calculate the distances between reversed sliding windows and every labeled sequence of page and action separately. A labeled service is recognized run-time if the edit distance is less than $1.5$.

 And the package-changing events can be monitored by Accessibility Event\cite{AccessibilityService} to recognize the opening of any applications.

\subsection{User Interface}
\subsubsection{Rule Extraction User Interface}
\begin{figure}[htbp]
    \centering
     \setlength{\belowcaptionskip}{-0.5cm}  
    \subfigure(a){
      \includegraphics[width=.45\linewidth]{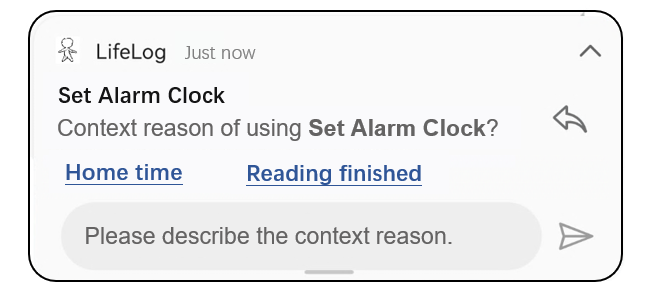}
    }
    \subfigure(b){
      \includegraphics[width=.4\linewidth]{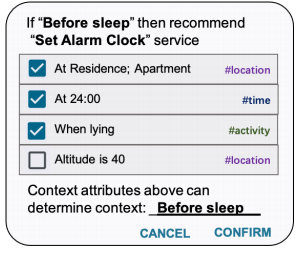}
    }
    \caption{(a) Notification to ask for reasons and to show the predicted context reasons(\textit{Home time}, \textit{Reading finished}). (b) List for users to select context attributes.}
    \label{fig:request & confirm}
\end{figure}

\paragraph{Notification to ask for interaction}
At the timing of in situ requests after service recognition, we use notification reminders to ask for reasons and show predicted reasons, as in Figure \ref{fig:request & confirm}(a). There is not always a contextual rule for the service, and users are not always wishing to extract. Notification reminders do not interrupt user's current interaction, with which the user can judge whether contextual rules need to be extracted.
\paragraph{List display to select context attributes}
Notice, this part of manual selection is solely designed to validate LLM's ability of correctly identifying essential context attributes. The result of the selection is only used for validation of the alignment between LLM's identification of essential attributes and users' intentions.
Users use the list selection method to choose the context attributes corresponding to the contextual rules. We attach colored tags to context attributes for users to find and identify them easily.
\subsubsection{Service Recommendation User Interface}
\begin{figure}[htbp]
    \centering
     \setlength{\belowcaptionskip}{-0.5cm}  
    \subfigure(a){
      \includegraphics[width=.4\linewidth]{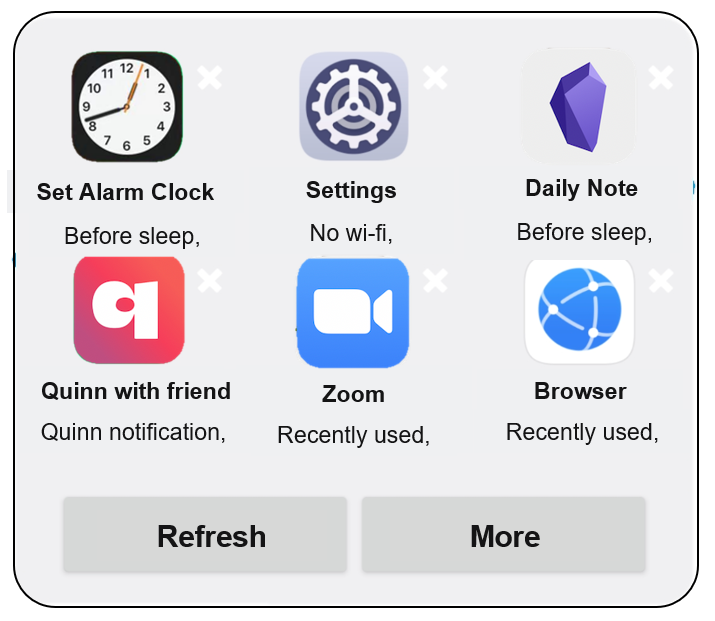}
    }
    \subfigure(b){
      \includegraphics[width=.45\linewidth]{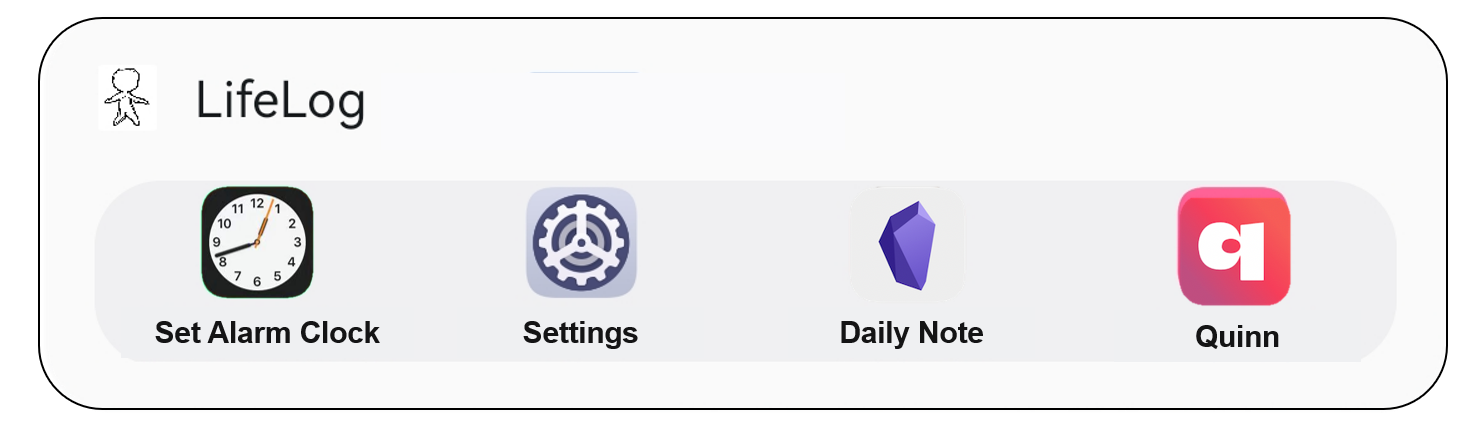}
    }
    \caption{(a) Home screen widget, showing the recommended services and reasons.  (b) Lock screen notification, showing the recommended services.}
    \label{fig:widget & lockscreen}
\end{figure}

\paragraph{Home screen widget}
SayRea implements a home screen widget that displays the recommended results, as shown in Figure \ref{fig:widget & lockscreen}(a). The icon, label, and recommendation reason are displayed for each service result, and clicking on them triggers the corresponding service. It also provides a "$\times$" delete button in the upper-right corner of the icon to remove the corresponding service recommendation result from the list and initiate an error feedback interaction.
The home screen widget is similar to the current application recommendation on smartphones, imposing low learning costs for users. In addition, recommendation reasons make SayRea more interpretable.
\paragraph{lock screen notifications}
 SayRea additionally implements a grid list notification, as shown in Figure \ref{fig:widget & lockscreen}(b). For each service result, only the icon and label of the service are displayed and clickable to trigger the service.
The widget still requires unlocking the phone, and the phone may not show the screen where the widget is located, lock screen notifications are relatively faster.

\section{User Study}
\par We conducted a 10-day field study with a total of 20 participants. We first asked each participant to sign an informed consent before the experiment. Notice that the university ethics review board approves human-subjects research and they approved this project. We then introduced SayRea to the participants and installed the application on their smartphones. We demonstrated how to say reasons for using particular services. We also ask the participants to choose corresponding context attributes for validation use. We required participants to place the widget that shows recommended services in an obvious place on the home screen. We also asked them to give necessary permissions to enable information gathering and internet connection of the application. 
\par Participants are asked to use SayRea naturally on their smartphones, without being required of any deliberate use. Throughout the 10-day field study, we logged users' behaviours and used objective metrics and subjective metrics to evaluate the system.
\subsection{Result}
\subsubsection{Accuracy of LLM's predictions}
\par We use GPT-3.5-turbo\cite{Brown2020Language} to predict context attributes based on semanticized contexts, services and users' reasons. A list of LLM's predicted context attributes is considered \textbf{accurate} when more than 75\% of users' choices of context attributes overlap with LLM's identification. In our field study, 85\% of LLM's predictions are accurate. This considerably high accuracy shows that LLM has a trustworthy ability to identify context attributes, subsequently to accumulate correct contextual rules.

\subsubsection{Accumulation of contextual rules}
\par
\begin{figure}[h]
  \centering
  \includegraphics[width=\linewidth]{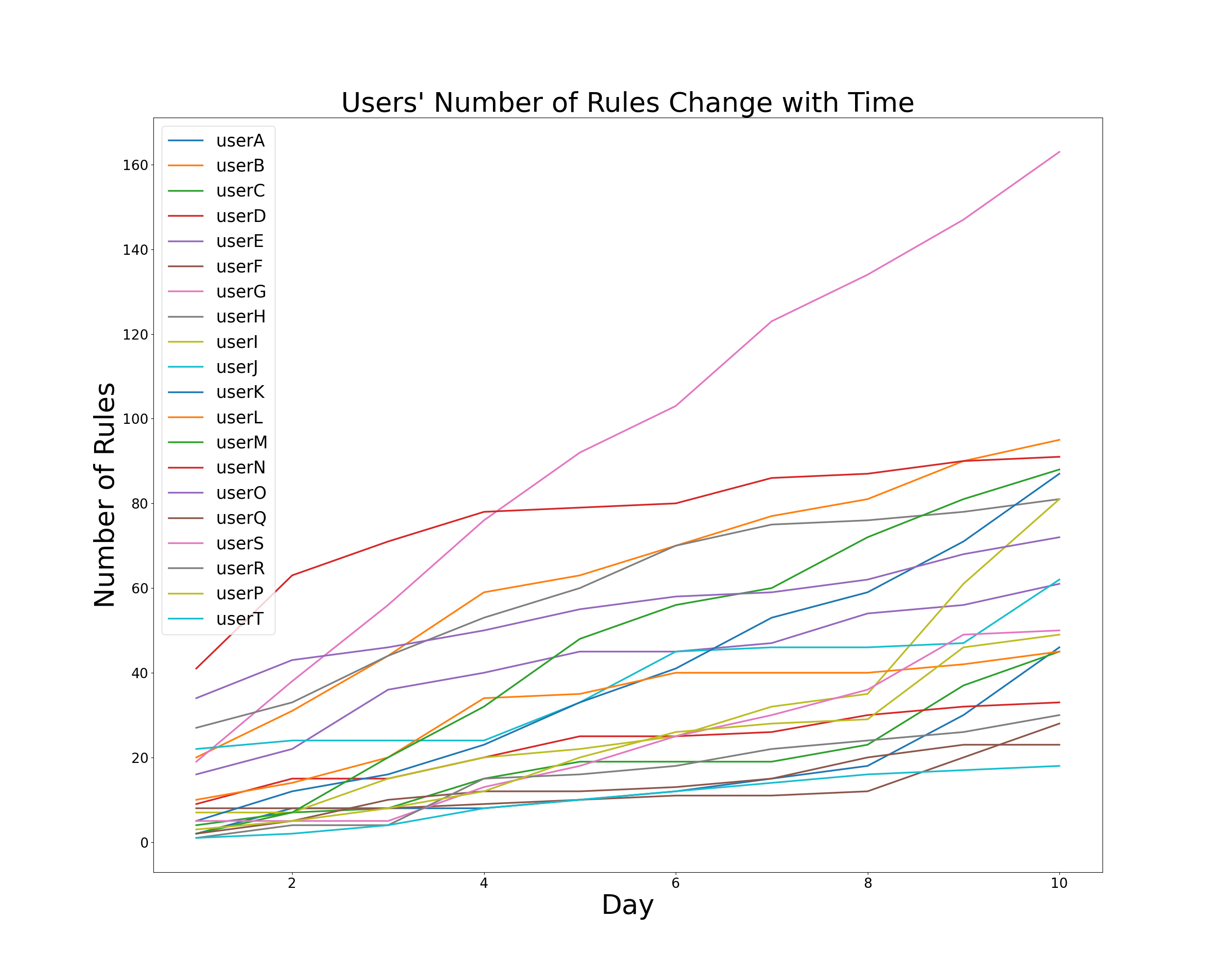}
  \caption{Accumulation of contextual rules of each user.}
  \label{fig:rules_num}
\end{figure}

 We calculated the number of contextual rules accumulated by each user. 50\% participants had accumulated more than 60 rules by the end of the experiment. 95\% participants accumulated more than 20 rules. One user accumulated 163 rules at the end of the experiment. Figure \ref{fig:rules_num} shows the accumulation of contextual rules of each user.
\par Results show that SayRea can accumulate a considerable number of contextual rules without participants' deliberate use of the system. Although the number of contextual rules accumulated may vary due to different smartphone using habits, it can be concluded that SayRea has shown the usability of the extraction method.

\subsubsection{Time consumption of rule extraction for users}
\par
\begin{figure}[h]
  \centering
  \includegraphics[width=\linewidth]{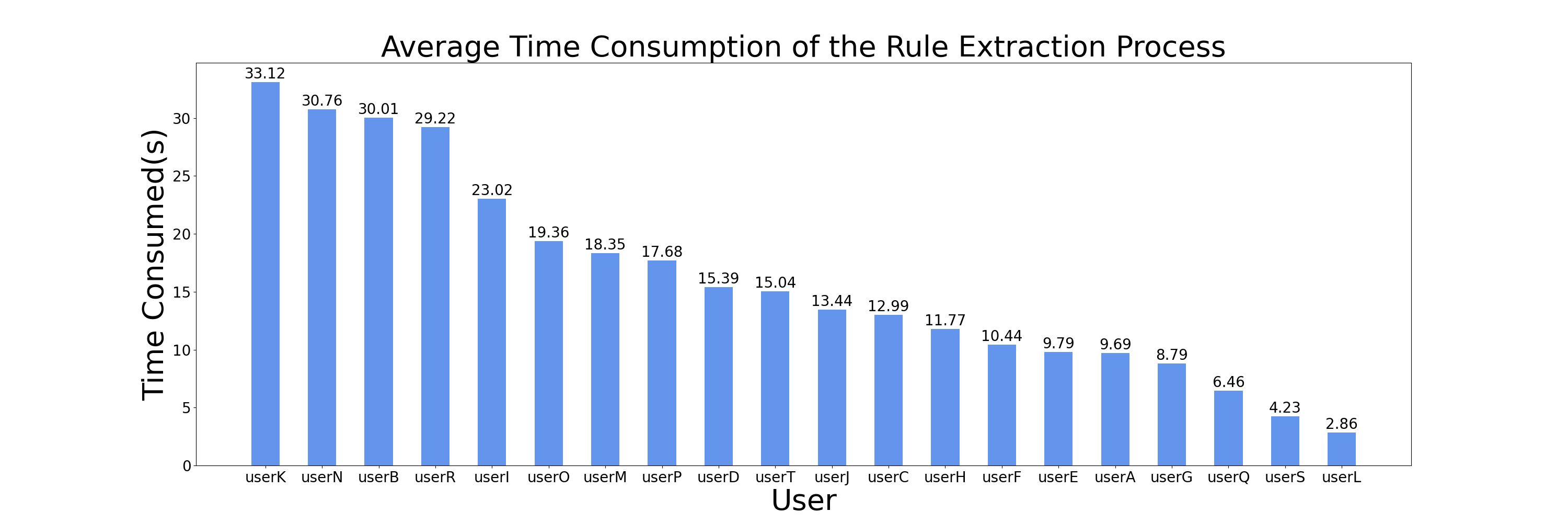}
  \caption{Time consumption of each user.}
  \label{fig:efficiency}
\end{figure}

 We calculated the time consumption of rule extraction(from "start reason input" to "finish context attribute selection") for each user. During this time, users can naturally continue their usage of the services when information(contexts, services and reasons) are being processed by the system. Results show that it takes a user 16.12 seconds to go through this process on average. 50\% participants spend less than 15 seconds to finish the extraction process. And the user who spends the least time only spends less than 3 seconds. UserS and UseA, who extract rules using predicted context reasons(Section \ref{sec:Accumulation and Optimization}) the most proportionally (20\%), spend significantly less time than the average time consumption. The considerably low time consumption of the rule extraction process shows that SayRea imposes low interaction burdens on users and is easy to use. Figure \ref{fig:efficiency} shows the time consumption for each user. It is worth noticing that this reported time include the contextual attributes selection process for validation purpose, which is not part of the approach.

\subsubsection{Coverage}
\par
\begin{figure}[h]
  \centering
  \includegraphics[width=\linewidth]{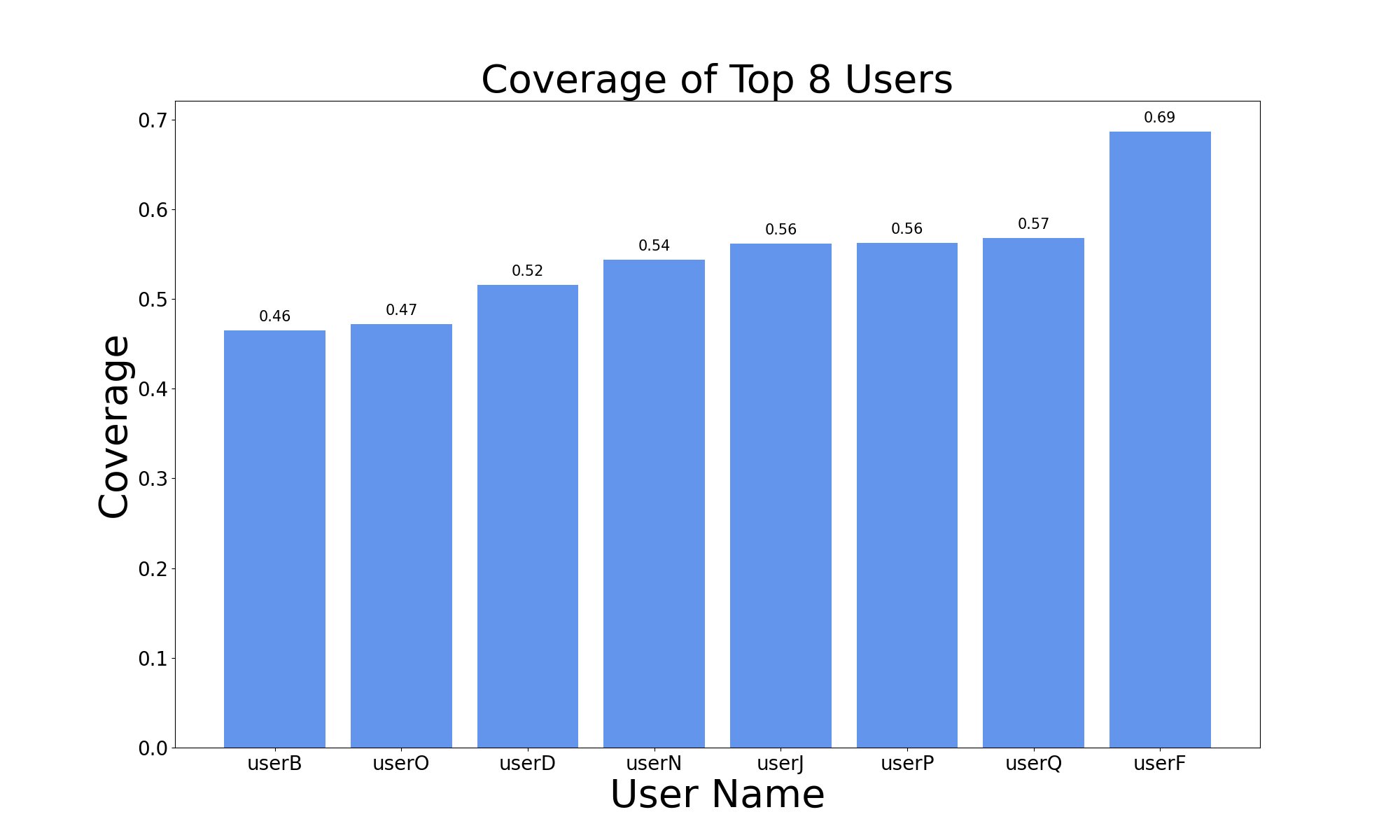}
  \caption{Top 8 coverage of users}
  \label{fig:coverage}
\end{figure}

 For each time a service is used, it is \textbf{covered} if SayRea is recommending it. 
$$ R_c = \frac{N_c}{N_a} ,$$where $R_c$ represents coverage, $N_c$ represents the times that services are covered, and $N_a$ represents the total times of service usage. Our system can achieve an average coverage of 45\% for all users at the end of the study. 30\% participants have an overall coverage of more than 50\% throughout the study. 95\% participants exceed an overall coverage of 20\%. One participant reaches an overall coverage of 69\%. The coverage's coefficient of variance is 0.32, indicating a relatively evenly distributed coverage. Figure \ref{fig:coverage} shows the top 8 coverage of all users.
\par
\begin{figure}[ht]
  \centering
  \includegraphics[width=\linewidth]{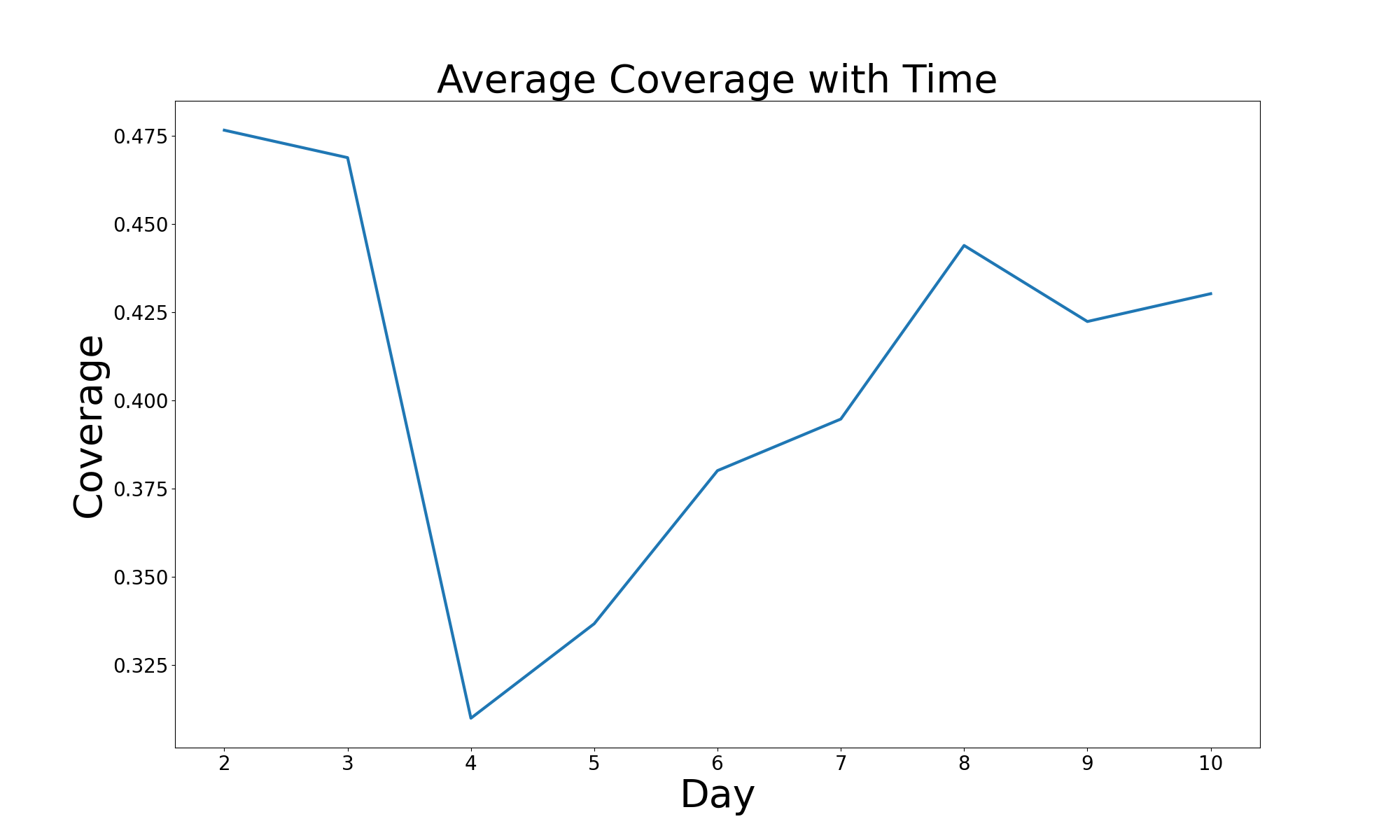}
  \caption{Change of average coverage with time.}
  \label{fig:coverage_time}
\end{figure}

 The average coverage shows a tendency to drop at the beginning of the study, and rise until being stable at around 45\%. SayRea recommends recently used services as its cold start strategy, accounting for the high coverage at the beginning because of the locality of service usage. Relatively limited rules after the cold start stage cause a drop in the coverage. With the use of our system, rules are accumulated, resulting in coverage's increase. Figure \ref{fig:coverage_time} shows how average coverage changes over time from day 2 to day 10. To better display the tendency, we omit day 1, whose data contains users' random exploration of SayRea. 
\par User study shows that the recommendations of SayRea can cover a significant part of users' service usage. Coverage can steadily increase over time before becoming stable. This shows our system's recommendation capability.

\subsubsection{Subjective Metrics}
\par The results of a 7-point Likert scale survey are displayed in Table 2. The results show that SayRea's usability, interpretability, and controllability have been recognized by most users, along with interactability, evolvability, and un-laboriousness.
\begin{table}[ht]
\centering
  \caption{Results of a likert scale survey.}
  \label{tab:Likert Scale Results}
  \begin{tabular}{p{5.8cm}c}
    \toprule
    \textbf{Statements} &\textbf{Results}\\
    \midrule
    You can tell the reason for the recommendation when a service is recommended. &5.11(sd=1.21) \\ \hline 
    You feel you have participated in building the system. &5.47(sd=1.09) \\ \hline 
    You feel the recommender system is recommending services based on your habits. &5.11(sd=1.16) \\ \hline 
    You can interact smoothly with the system. &5.21(sd=1.24) \\ \hline 
    You feel the system's ability is evolving. &5.42(sd=1.18) \\ \hline 
    You feel it is not hard to use the system. &5.84(sd=0.93) \\
  \bottomrule
\end{tabular}
\end{table}
\par We further discussed with participants why they find the system not hard to use (5.84 in 7, as shown in Table \ref{tab:Likert Scale Results}). Most thoughts point out that they feel SayRea can actually "understand" them with merely a reason. All they need to do is to say a simple reason without heavy burdens of thinking. Many participants also pointed out that SayRea was the first recommender system they had used that took users' own reasons into consideration, making them feel that they were in control of the system which they were building by themselves. This corresponds to the second statement shown in Table \ref{tab:Likert Scale Results}.

\section{Discussion}
\subsection{About Coverage}
\par Although SayRea reaches an average of 45\% coverage and a maximum of almost 70\% coverage, it can be argued that there is room for improvement. We believe there are mainly two reasons for the improvable coverage.
\begin{itemize}
    \item Users' usage of smartphones cannot be totally explained by context reasons\cite{ChallengesCARS}. We discussed this with several users who have relatively low coverage. Their reasons for the low coverage can be summarized as they just did not think their usage of service or their reasons have any relationship with the contexts taken into consideration, resulting in a weak or meaningless rule. SayRea now supports relatively limited types of contexts, which may account for the insufficient candidate context attributes. But some services, like opening a Camera to "take photos of a beautiful butterfly", just cannot be mapped to any context attribute according to interviewed participants.
    \item Users sometimes are not aware of the casual rules. We discuss in Limitation (Section \ref{sec:Automatic Discovery of Contextual Rules}) that the automatic discovery of new contextual rules can extract contextual rules without deliberate labeling, which can help users find unaware contextual rules to improve coverage.
\end{itemize}

\subsection{Lack of Comparisons with Baselines}
In addition to current context attributes and services, SayRea recommends based on users' own reasons, which are highly personalized and changeable. There lacks recommender system baselines that take users' own reasons into consideration. And even such baselines exist, results of our field study cannot be easily compared with those of laboratary experiments in order to demonstrate SayRea's performance. Further, existing recommender system baselines are difficult to run and evaluate\cite{rendle2019difficulty}, making comparable baselines for SayRea hard to exist. Our main novelty is that our approach is low-burden and in-situ, the considerably good results as shown in the user study further reduce the necessity of the comparisons with baselines.

\section{Limitations}
\subsection{Automatic Discovery of Contextual Rules}
\label{sec:Automatic Discovery of Contextual Rules}
SayRea extracts a contextual rule when it recognizes a service, but it is not always the case that the service is a result of a contextual rule. So the frequent request, even with a designed notification reminder, can still cause disturbance to users. Future systems can use sophisticated algorithms to evaluate the possibility of contextual rules and interact with users when such possibility is high. Moreover, they can incorporate automatic contextual rule discovery algorithms like association rule mining \cite{Iqbal2020ABC-RuleMiner} to omit non-essential interactions and reduce the interaction burden. However, it is worth noticing that the above methods may require data-driven algorithms and numerous training corpus. The method proposed in this paper is a contextual rule extraction algorithm with high availability for the cold start when the amount of data is low.

\subsection{Context Extension}
In this paper, SayRea supports eight types of contexts. They do not fully describe the contexts of the service usage, which can confuse the user when selecting context attributes. However, SayRea has good contextual extensibility. Any new context can be converted into context attributes by context semanticization, and different context attributes can be understood with semantic information through LLM. Some works allow users to teach the intelligent system certain context acquisitions by PBD (program by demonstration), such as PUMICE \cite{toby2019PUMICE}. SayRea can refer to these works to support interactively defined contexts, making full use of the user's willingness to interact,

\subsection{Further Management of Context Attributes}
SayRea uses a context rule tree to manage contextual rules, in which all context attributes are considered homogeneous. However, various types of relationships may exist between different types and values of context attributes, such as mutual exclusion, intersection, inclusion, and approximation \cite{Karchoud2019One}. These relationships can be utilized to enhance the system's understanding of the context. Context attributes can also be combined in other forms (e.g., logical or) for context causes in contextual rules (which can only be handled as different contextual rules in the current system), enriching the causes made of context attributes to express more complex contexts.

\section{Conclusion}
In this paper, we propose an in situ interactive method for extracting contextual rules. This contextual rule extraction method enables the low-burden user participation at the appropriate times (in situ) to extract contextual relationships between contexts and services with Large Language Models. It has the potential to be extended to other recommendation or contextual rule extraction fields.

SayRea, a context-based service recommender system with the contextual rule extraction method, has good usability, interpretability, and controllability. We implemented SayRea on Android and conducted field experiments. SayRea performed well in terms of rule accumulation and service coverage, indicating the high usability of the system. Subjective feedback from users showed good acceptance of SayRea, and users perceived the system as having well explainability and controllability.

\bibliographystyle{unsrt}  
\bibliography{references}

\begin{thebibliography}{10}

\bibitem{Ma2020Roaming}
Yun Ma, Ziniu Hu, Diandian Gu, Li~Zhou, Qiaozhu Mei, Gang Huang, and Xuanzhe Liu.
\newblock Roaming through the castle tunnels: An empirical analysis of inter-app navigation of android apps.
\newblock {\em ACM Trans. Web}, 14(3), jun 2020.

\bibitem{Steven2014IFTTT}
Steven Ovadia.
\newblock Automate the internet with “if this then that” (ifttt).
\newblock {\em Behavioral \& Social Sciences Librarian}, 33(4):208--211, 2014.

\bibitem{Ilarri2022An}
Sergio Ilarri, Irene Fumanal, and Raquel Trillo-Lado.
\newblock An experience with the implementation of a rule-based triggering recommendation approach for mobile devices.
\newblock In {\em The 23rd International Conference on Information Integration and Web Intelligence}, iiWAS2021, page 562–570, New York, NY, USA, 2022. Association for Computing Machinery.

\bibitem{Brown2020Language}
Tom~B. Brown, Benjamin Mann, Nick Ryder, Melanie Subbiah, Jared Kaplan, Prafulla Dhariwal, Arvind Neelakantan, Pranav Shyam, Girish Sastry, Amanda Askell, Sandhini Agarwal, Ariel Herbert-Voss, Gretchen Krueger, Tom Henighan, Rewon Child, Aditya Ramesh, Daniel~M. Ziegler, Jeffrey Wu, Clemens Winter, Christopher Hesse, Mark Chen, Eric Sigler, Mateusz Litwin, Scott Gray, Benjamin Chess, Jack Clark, Christopher Berner, Sam McCandlish, Alec Radford, Ilya Sutskever, and Dario Amodei.
\newblock Language models are few-shot learners.
\newblock In {\em Proceedings of the 34th International Conference on Neural Information Processing Systems}, NIPS'20, Red Hook, NY, USA, 2020. Curran Associates Inc.

\bibitem{Musumba2013Context-awareness}
George~W. Musumba and Henry~O. Nyongesa.
\newblock Context awareness in mobile computing: a review.
\newblock {\em International Journal of Machine Learning and Application}, 2:1--10, Jan 2013.

\bibitem{María2021context-aware}
María {del Carmen Rodríguez-Hernández} and Sergio Ilarri.
\newblock Ai-based mobile context-aware recommender systems from an information management perspective: Progress and directions.
\newblock {\em Knowledge-Based Systems}, 215:106740, 2021.

\bibitem{Norha2018Characterizing}
Norha~M. Villegas, Cristian Sánchez, Javier Díaz-Cely, and Gabriel Tamura.
\newblock Characterizing context-aware recommender systems: A systematic literature review.
\newblock {\em Knowledge-Based Systems}, 140:173--200, 2018.

\bibitem{Carrasco2019Siri}
Manuel Carrasco~Molina.
\newblock {\em Siri and Search}, pages 139--188.
\newblock Apress, Berkeley, CA, 2019.

\bibitem{shen2019deepapp}
Zhihao Shen, Kang Yang, Wan Du, Xi~Zhao, and Jianhua Zou.
\newblock Deepapp: A deep reinforcement learning framework for mobile application usage prediction.
\newblock In {\em Proceedings of the 17th Conference on Embedded Networked Sensor Systems}, SenSys '19, page 153–165, New York, NY, USA, 2019. Association for Computing Machinery.

\bibitem{Zhu2021CMARA}
Ke~Zhu, Yingyuan Xiao, Wenguang Zheng, Xu~Jiao, and Ching-Hsien Hsu.
\newblock A novel context-aware mobile application recommendation approach based on users behavior trajectories.
\newblock {\em IEEE Access}, 9:1362--1375, 2021.

\bibitem{Meng2022Survey}
X.~Meng, Y.~Du, Y.~Zhang, and X.~Han.
\newblock A survey of context-aware recommender systems: From an evaluation perspective.
\newblock {\em IEEE Transactions on Knowledge; Data Engineering}, 0(01):1--20, jun 2022.

\bibitem{Yongfeng2020Explainable}
Yongfeng Zhang and Xu~Chen.
\newblock Explainable recommendation: A survey and new perspectives.
\newblock {\em Foundations and Trends® in Information Retrieval}, 14(1):1--101, 2020.

\bibitem{Tsai2021controllability}
Chun-Hua Tsai and Peter Brusilovsky.
\newblock The effects of controllability and explainability in a social recommender system.
\newblock {\em User Modeling and User-Adapted Interaction}, 31(3):591--627, Jul 2021.

\bibitem{Jannach2017Control}
Dietmar Jannach, Sidra Naveed, and Michael Jugovac.
\newblock User control in recommender systems: Overview and interaction challenges.
\newblock In Derek Bridge and Heiner Stuckenschmidt, editors, {\em E-Commerce and Web Technologies}, pages 21--33, Cham, 2017. Springer International Publishing.

\bibitem{Scheel2014Explanations}
Christian Scheel, Angel Castellanos, Thebin Lee, and Ernesto~William De~Luca.
\newblock The reason why: A survey of explanations for recommender systems.
\newblock In Andreas N{\"u}rnberger, Sebastian Stober, Birger Larsen, and Marcin Detyniecki, editors, {\em Adaptive Multimedia Retrieval: Semantics, Context, and Adaptation}, pages 67--84, Cham, 2014. Springer International Publishing.

\bibitem{Amit2022Evolving}
Amit Livne, Eliad~Shem Tov, Adir Solomon, Achiya Elyasaf, Bracha Shapira, and Lior Rokach.
\newblock Evolving context-aware recommender systems with users in mind.
\newblock {\em Expert Systems with Applications}, 189:116042, 2022.

\bibitem{Zhang2019Context-Aware}
Guoming Zhang, Lianyong Qi, Xuyun Zhang, Xiaolong Xu, and Wanchun Dou.
\newblock Context-aware point-of-interest recommendation algorithm with interpretability.
\newblock In Xinheng Wang, Honghao Gao, Muddesar Iqbal, and Geyong Min, editors, {\em Collaborative Computing: Networking, Applications and Worksharing}, pages 745--759, Cham, 2019. Springer International Publishing.

\bibitem{Li2021CAESAR}
Lei Li, Li~Chen, and Ruihai Dong.
\newblock Caesar: context-aware explanation based on supervised attention for service recommendations.
\newblock {\em Journal of Intelligent Information Systems}, 57(1):147--170, Aug 2021.

\bibitem{Karchoud2017Long-life}
Riadh Karchoud, Arantza Illarramendi, Sergio Ilarri, Philippe Roose, and Marc Dalmau.
\newblock Long-life application.
\newblock {\em Personal and Ubiquitous Computing}, 21(6):1025--1037, Dec 2017.

\bibitem{Zhang2002Causality}
Chengqi Zhang and Shichao Zhang, editors.
\newblock {\em Causality in Databases}, pages 85--120.
\newblock Springer Berlin Heidelberg, Berlin, Heidelberg, 2002.

\bibitem{Shaina2019overview}
Shaina Raza and Chen Ding.
\newblock Progress in context-aware recommender systems — an overview.
\newblock {\em Computer Science Review}, 31:84--97, 2019.

\bibitem{Iqbal2020ABC-RuleMiner}
Iqbal~H. Sarker and A.S.M. Kayes.
\newblock Abc-ruleminer: User behavioral rule-based machine learning method for context-aware intelligent services.
\newblock {\em Journal of Network and Computer Applications}, 168:102762, 2020.

\bibitem{Zaki2004Mining}
Mohammed~J. Zaki.
\newblock Mining non-redundant association rules.
\newblock {\em Data Mining and Knowledge Discovery}, 9(3):223--248, Nov 2004.

\bibitem{Coronado2016Automation}
Miguel Coronado and Carlos~A. Iglesias.
\newblock Task automation services: Automation for the masses.
\newblock {\em IEEE Internet Computing}, 20(1):52--58, Jan 2016.

\bibitem{Mi2017IFTTT}
Xianghang Mi, Feng Qian, Ying Zhang, and XiaoFeng Wang.
\newblock An empirical characterization of ifttt: Ecosystem, usage, and performance.
\newblock In {\em Proceedings of the 2017 Internet Measurement Conference}, IMC '17, page 398–404, New York, NY, USA, 2017. Association for Computing Machinery.

\bibitem{Shortcutonline}
Apple.
\newblock Shortcuts user guide – apple support, 2023.

\bibitem{Karchoud2019One}
Riadh Karchoud, Philippe Roose, Marc Dalmau, Arantza Illarramendi, and Sergio Ilarri.
\newblock One app to rule them all: collaborative injection of situations in an adaptable context-aware application.
\newblock {\em Journal of Ambient Intelligence and Humanized Computing}, 10(12):4679--4692, Dec 2019.

\bibitem{Boyaci2011Bridging}
Omer Boyaci, Victoria Beltran, and Henning Schulzrinne.
\newblock Bridging communications and the physical world: Sense everything, control everything.
\newblock In {\em Proceedings of the 5th International Conference on Principles, Systems and Applications of IP Telecommunications}, IPTcomm '11, New York, NY, USA, 2011. Association for Computing Machinery.

\bibitem{Chihani2013DOTT}
Bachir Chihani, Emmanuel Bertin, and Noel Crespi.
\newblock {A user-centric context-aware mobile assistant}.
\newblock In {\em {ICIN 2013 : 17th International Conference on Intelligence in Next Generation Networks}}, pages 110 -- 117, Venice, Italy, October 2013. {IEEE Computer Society}.

\bibitem{Seyff2014AppEcho}
Norbert Seyff, Gregor Ollmann, and Manfred Bortenschlager.
\newblock Appecho: A user-driven, in situ feedback approach for mobile platforms and applications.
\newblock In {\em Proceedings of the 1st International Conference on Mobile Software Engineering and Systems}, MOBILESoft 2014, page 99–108, New York, NY, USA, 2014. Association for Computing Machinery.

\bibitem{korn2013potentials}
Oliver Korn, Albrecht Schmidt, and Thomas H{\"o}rz.
\newblock The potentials of in-situ-projection for augmented workplaces in production: a study with impaired persons.
\newblock In {\em CHI'13 Extended Abstracts on Human Factors in Computing Systems}, pages 979--984. 2013.

\bibitem{hoffswell2018augmenting}
Jane Hoffswell, Arvind Satyanarayan, and Jeffrey Heer.
\newblock Augmenting code with in situ visualizations to aid program understanding.
\newblock In {\em Proceedings of the 2018 CHI Conference on Human Factors in Computing Systems}, pages 1--12, 2018.

\bibitem{cantino2021prompt}
Andrew Cantino.
\newblock Prompt engineering tips and tricks with gpt-3, 2021.

\bibitem{Wu2020Tree}
Zhiang Wu, Changsheng Li, Jie Cao, and Yong Ge.
\newblock On scalability of association-rule-based recommendation: A unified distributed-computing framework.
\newblock {\em ACM Trans. Web}, 14(3), jun 2020.

\bibitem{ContentDescription}
Android.
\newblock Make apps more accessible | android developers, 2023.

\bibitem{AccessibilityService}
Android.
\newblock Accessibilityservice|androiddevelopers, 2023.

\bibitem{Levenshtein}
V.~I. {Levenshtein}.
\newblock {Binary Codes Capable of Correcting Deletions, Insertions and Reversals}.
\newblock {\em Soviet Physics Doklady}, 10:707, February 1966.

\bibitem{ChallengesCARS}
Zhang Yujie and Wang Licai.
\newblock Some challenges for context-aware recommender systems.
\newblock In {\em 2010 5th International Conference on Computer Science \& Education}, pages 362--365, -, 2010. IEEE, -.

\bibitem{rendle2019difficulty}
Steffen Rendle, Li~Zhang, and Yehuda Koren.
\newblock On the difficulty of evaluating baselines: A study on recommender systems, 2019.

\bibitem{toby2019PUMICE}
Toby Jia-Jun Li, Marissa Radensky, Justin Jia, Kirielle Singarajah, Tom~M. Mitchell, and Brad~A. Myers.
\newblock Pumice: A multi-modal agent that learns concepts and conditionals from natural language and demonstrations.
\newblock In {\em Proceedings of the 32nd Annual ACM Symposium on User Interface Software and Technology}, UIST '19, page 577–589, New York, NY, USA, 2019. Association for Computing Machinery.

\end{thebibliography}

\end{document}